\documentclass[aps,prd,amssymb,nobibnotes, amsmath,twocolumn,nofootinbib,superscriptaddress,floatfix,preprintnumbers]{revtex4-1}

\usepackage{graphicx,color}% Include figure files
\usepackage{dcolumn}% Align table columns on decimal point
\usepackage{bm}% bold math
\usepackage{hyperref}% add hypertext capabilities
\usepackage{amsmath}
\usepackage{mathtools}
\usepackage{multirow}
\usepackage{stackrel}
\usepackage[lofdepth,lotdepth,caption=false]{subfig}
\usepackage{tikz-feynman}
\usepackage{verbatim}
\usepackage{xcolor}
\def\d{{\rm d}}

\usepackage{float}

\begin{document}

%===========================================================

\begin{flushleft}
LAPTH-053/24
\end{flushleft}

\title{Pair production with capture by energetic cosmic ray nuclei in a photon background}

\author{AmirFarzan Esmaeili}
\email{a.farzan.1993@aluno.puc-rio.br}
\affiliation{Departamento de Física, Pontifícia Universidade Católica do Rio de Janeiro, Rio de Janeiro 22452-970, Brazil}

\author{Arman Esmaili}
\email{arman@puc-rio.br}
\affiliation{Departamento de Física, Pontifícia Universidade Católica do Rio de Janeiro, Rio de Janeiro 22452-970, Brazil}

\author{Pasquale Dario~Serpico}
\email{serpico@lapth.cnrs.fr}
\affiliation{LAPTh, CNRS, USMB, F-74940 Annecy, France}

\begin{abstract}

We investigate the ionization state of very energetic cosmic ray nuclei in photon fields, such as the cosmic microwave background (CMB) in extragalactic propagation and the environment surrounding  the acceleration site in astrophysical sources. We focus on the process of pair production with electron capture (PPC), where the interaction of a single photon with a nucleus produces an $e^\pm$ pair (similar to Bethe-Heitler process) with the subsequent capture of electron by the nucleus. This process effectively reduces the nucleus charge by one unit and counteracts the photo-ionization process. We show that during cosmological propagation, where the ultra-high energy cosmic rays interact predominantly with the CMB, ionization dominates over PPC for all the cases of practical interest. However, within the source environment and at sufficiently high energies, ionization and PPC processes can reach an equilibrium, leading to a significant fraction of dressed heavy nuclei. This provides a further limitation to the acceleration of high-$Z$ nuclei to very high energies in environments with hot and dense photon fields.

\end{abstract}
\maketitle

%%%%%%%%%%%%%%%%%%%%%%%%%%%%%%%%%%
%%%%%%%%%%%%%%%%%%%%%%%%%%%%%%%%%%
\section{Introduction}
%%%%%%%%%%%%%%%%%%%%%%%%%%%%%%%%%%
%%%%%%%%%%%%%%%%%%%%%%%%%%%%%%%%%%

The enigmatic nature of ultra-high energy cosmic rays (UHECRs) has fueled decades of research, yet remain unresolved~\cite{Hillas1984,RevModPhys.72.689,Kotera_2011,AlvesBatista:2019tlv}. Observations from the Pierre Auger Observatory (PAO)~\cite{PierreAuger:2015eyc} and the Telescope Array (TA)~\cite{TelescopeArrayFD_2012,TelescopeArraySD_2012} have revealed key features of the UHECR energy spectrum. One such feature is the {\it ankle}, a distinct spectral hardening observed at around $\sim10^{18.7}$~eV. Another is the {\it instep} at $\sim 10^{19.1}$~eV, which may indicate the transition between light and intermediate nuclei~\cite{PAOUHECR2020}. Finally, a suppression in the total flux above $\sim10^{19.7}$~eV which can be due to energy losses during UHECR propagation~\cite{GZK1966,Zatsepin:1966jv}, the upper energy limits of particle acceleration at the sources, or a combination of both. The energy spectrum measured by TA aligns with PAO's results in both shape and normalization, within systematic uncertainties. However, some discrepancies highlight the inherent challenges in UHECR acceleration and propagation~\cite{TelescopeArray:2024tbi,Tsunesada:2021qO,Tsunesada:2023c0}; {\it e.g.}, the excess of events at the highest energies in TA, specially with the recent detection of the {\it Amaterasu} particle with the energy $\sim244$~EeV~\cite{Amaterasu:2023sbd}.

The composition of UHECRs can provide key information for the identification  of their astrophysical origins~\cite{PierreAuger:2022atd} and is closely related to their propagation in the galactic and intergalactic media. Several observations point toward the ubiquity of heavy nuclei in UHECRs, especially at higher energies. Data from PAO and TA indicate a mixed composition, with intermediate-mass nuclei ({\it e.g.}, carbon and oxygen) and heavy nuclei ({\it e.g.}, iron) becoming dominant above 10~EeV, while the proton fraction decreases at higher energies. Notably, the presence of ultra-heavy elements (i.e., with atomic number $Z\gg 26$) has been proposed as a potential source for the “second knee” in the UHECR spectrum~\cite{Lv:2024wrs}. 
At the highest energies, the production and acceleration of ultra-heavy UHECRs have also been  studied in the literature, particularly in the context of neutron-rich astrophysical phenomena such as collapsars~\cite{Murase:2008mr, Wang:2007xj, Murase:2010va, Metzger_2011, 2012ApJ...746...40L, Horiuchi_2012,Zhang:2017moz,Boncioli:2018lrv,Bhattacharya:2021cjc} and compact binary mergers involving neutron stars~\cite{Takami:2013rza, Kyutoku:2016ckb,Kimura:2018ggg,Rodrigues:2018bjg}. Recent works have also examined the cosmological propagation of ultra-heavy UHECRs~\cite{Zhang:2024sjp}. 

One of the main motivations for the recent interest in ultra-heavy UHECRs is that they impose significantly looser conditions on the acceleration to the highest energies. Almost always, a heavy nucleus ${}_Z^AX$, of mass number $A$ and atomic number $Z\lesssim A/2$, is assumed to be fully ionized with its charge equal to $Ze$; making abstraction of species-dependent energy-loss mechanisms, a given accelerator capable of an effective potential $\Delta V$ allows for a maximum energy $E_{\rm max}=Ze \Delta V$, proportionally higher for a nucleus than a proton. The charge of the UHECRs plays role also in their propagation after escaping the source via charge-dependent deflection and energy degradation in the galactic and/or extragalactic magnetic fields, making UHECR astronomy conversely more difficult with nuclei. 

However, this picture changes if the heavy nucleus ``dresses'' with one or more electrons, effectively reducing its charge. To our knowledge, this possibility has been overlooked in the literature since, at low energies, photo-ionization rates are orders of magnitude faster than any other process. Reassessing the extent to which this conclusion stays true at ultra-high energies for (ultra-)heavy nuclei is the main subject of this article.

The dressing process involves the interaction of ultra-heavy UHECRs with photons. In the course of propagation over cosmological distances, CRs encounter low-energy photon fields permeating the Universe, with the CMB being the most prominent one. Additionally, in the scenarios where CRs are accelerated inside compact astrophysical sources, such as the magnetosphere of a Super Massive Black Hole (SMBH) or Gamma Ray Bursts (GRBs), the acceleration site contains dense background photon fields. In fact, recent multi-messenger searches for the origin of high energy neutrinos observed by IceCube~\cite{IceCube:2020wum,IceCube:2023sov}, in combination with diffuse gamma-ray data~\cite{Fermi-LAT:2014ryh}, strongly point to opaque sources which contain relatively dense photon fields~\cite{Capanema:2020oet,Capanema:2020rjj,Murase:2013rfa,Fang:2022trf}. Then, the observation of neutrinos with energy up to $\sim10^{16}$~eV by IceCube suggests then sources surrounded by dense photon fields exist that may accelerate protons (nuclei) up to $\sim10^{17}\div10^{18}$~eV ($\gtrsim10^{19}$~eV). Sufficiently energetic photons interacting with such nuclei can produce an $e^\pm$ pair (Bethe-Heitler process), with a non-vanishing probability that the electron is captured, thereby reducing the effective charge of the nucleus (or ion) by one unit. Of course, this process is unavoidably accompanied by its inverse, where the same target photons can ionize a dressed nucleus and increase its charge by one unit. These processes, known as pair production with capture (PPC) and photo-ionization, respectively, are the focus of this study. Both the PPC and photo-ionization can play important role for ultra-heavy UHECRs due to their scaling $\propto Z^5$. The ion can in principle reach a state of equilibrium for some effective charge, denoted as $Z_{\rm eq}$, lower than the nucleus atomic number, such that the PPC rate at $Z_{\rm eq}$ is equal to photo-ionization rate of the ion with charge $(Z_{\rm eq}-1)$. Of course, the dressing is relevant for a significant fraction of nuclei at some energy $E$ only if the PPC rate is sufficiently fast, compared to the timescale with which substantial energy loss due to the well-known mechanisms (such as Bethe-Heitler pair production, photo-meson production or photo-disintegration) take place.   

We note that the possibility of an equilibrium between PPC and photo-ionization during the cosmological propagation of heavy nuclei has been briefly mentioned (with different motivations) in~\cite{Tinyakov:2006jm}. Due to their approximations, we disagree with their conclusions, but for the (rather academic) case of protons.

This article is structured as follows: in Section~\ref{sec:rate} we introduce the key ingredients used to evaluate the rates of PPC and photo-ionization processes. To determine the time evolution of the ionization state and the possible equilibrium condition, we discuss and solve a set of Saha-like equations, described in Section~\ref{sec:saha}. Our main results are presented in Section~\ref{sec:saha}. Finally, Section~\ref{sec:conclusion} includes a discussion, possible directions for future investigations and applications, and our conclusions. 

%%%%%%%%%%%%%%%%%%%%%%%%%%%%%%%%%%
%%%%%%%%%%%%%%%%%%%%%%%%%%%%%%%%%%
\section{PPC and Photo-Ionization Rates}\label{sec:rate}
%%%%%%%%%%%%%%%%%%%%%%%%%%%%%%%%%%
%%%%%%%%%%%%%%%%%%%%%%%%%%%%%%%%%%

The energy spectra of nuclei emitted from astrophysical sources are shaped by the complex interplay between acceleration, cooling, and escape processes (see e.g. \cite{Murase:2019vdl}). As these nuclei propagate over cosmological distances, their spectra are further modified by energy loss processes, primarily through interactions with background photon fields such as the CMB. 
By introducing the species ionization state as further dynamical variable, as advocated here, one greatly complexifies the treatment: In principle, by considering nuclei up to a mass number $A_{\rm max} $, instead of dealing with $\sim A_{\rm max}$ different  kinetic (or stochastic) differential equations, one should now solve $\sum_i A_i\times Z_i\propto A_{\rm max}^2$ equations, in order to describe abundances, spectra, and ionization states of each species, with an interplay among those variables. In practice, and in particular for a first assessment of the relevance of the effect, {\it if the PPC and ionization interaction rates exceed the energy-loss rates,  we can factorize the problem and separately compute the typical ionization state  mixture of nuclei having a certain energy, as attained over a length scale set by the energy losses}. This approximation fails when this ``fast PPC'' hypothesis is not met. However, in this regime we expect small effects, so one has limited interest in solving for them.

To evaluate the impact of PPC and photo-ionization processes, the key parameters are the interaction and loss rates or, equivalently, the interaction and loss lengths. For a nucleus with energy $E_A$ and mass $m_A$ interacting with a photon field of spectrum $n(\varepsilon_\gamma)$ in the Lab frame, the {\it interaction rate} for a process $p$ is given by
\begin{equation}\label{eq:int_rate}
    \Gamma_{p}^{\rm int}(E_A) = \int \d\varepsilon_{\gamma} \int_{-1}^{1}\d\mu  \,P(\mu)n(\varepsilon_\gamma)\sigma_p(s)\,, 
\end{equation}
where $\sigma_p$ denotes the total cross section of the process, $\mu$ represents the cosine of the collision angle in the photon-nucleus interaction, $s = m_A^2 + 2\varepsilon_\gamma(E_A - \mu\sqrt{E_A^2 - m_A^2})$ is the center of momentum (CoM) energy squared, and $P(\mu) = (1-\beta_A\mu)/2$ is the flux factor, with $\beta_A$ being the velocity of the nucleus in the Lab frame. In a similar way, we define the {\it loss rate} as
\begin{equation}\label{eq:loss_rate}
    \Gamma_{p}^{\rm loss}(E_N) = \int \d\varepsilon_{\gamma} \int_{-1}^{1}\d\mu  \,P(\mu)n(\varepsilon_\gamma)\sigma_p(s)\eta(s)\,, 
\end{equation}
where $\eta(s)$ is the inelasticity of the interaction.

In section~\ref{subsec:PPC_ION} we review the total cross sections of PPC and photo-ionization. Using these established cross sections in Eq.~\eqref{eq:int_rate}, we calculate the corresponding rates in two distinct scenarios: interaction with the CMB during cosmological propagation and inside astrophysical sources where a thermal photon field is present. In section~\ref{subsec:BH_PD} we discuss Bethe-Heitler, photo-disintegration and photo-meson processes as the most relevant energy loss mechanisms for nuclei in our context. We then provide the associated loss and interaction lengths. Note that, unless stated otherwise, we are using natural units in our equations. Also, quantities in the nucleus rest frame are distinguished by prime ($'$) notation. 

%%%%%%%%%%%%%%%%%%%%%%%%%%%%%%%%%%
%%%%%%%%%%%%%%%%%%%%%%%%%%%%%%%%%%
\subsection{Total Cross Sections and Rates of PPC and Photo-Ionization}\label{subsec:PPC_ION}
%%%%%%%%%%%%%%%%%%%%%%%%%%%%%%%%%%
%%%%%%%%%%%%%%%%%%%%%%%%%%%%%%%%%%

The capture and ionization processes are most pronounced for the atomic $K$-shell of the nucleus, so we focus on the relevant formulae specific to the $K$-shell. The cross section of PPC in the $K$-shell can be derived from its inverse process, where a positron annihilates with a $K$-shell electron, resulting in photon emission, as described in~\cite{PhysRev.135.A1232}. In the Born approximation limit ($\alpha Z \ll 1$) this cross section is given by
\begin{equation}\label{eq:XS_PPC}
    \sigma_{\rm PPC}(E'_+)=\frac{4\pi \alpha^6 Z^5}{m_e\sqrt{E'^{2}_+-m_e^2}}\frac{E'^2_+ -\frac{m_e E'_+}{3}+\frac{m_e^2}{3}}{(E'_+ + m_e)^2}\,.
\end{equation}
Here, $E'_+$ denotes the outgoing positron energy, $\alpha$ and $m_e$ correspond to the fine structure constant and the electron mass, respectively. In the nucleus rest frame, the incident photon energy, $\varepsilon'_\gamma$, the ionization energy, $I_K$, and the positron energy are related by
\begin{equation}\label{eq:E+CoM}
    \varepsilon'_\gamma = E'_+ +m_e - I_K\,,
\end{equation}
where the $K$-shell ionization energy writes $I_K = m_e(\alpha Z)^2/2$. The $K$-shell photo-ionization cross section in the relativistic regime, applicable when $\alpha Z \ll 1$, calculated in~\cite{Akhiezer:1986yqm}, is given by
\begin{equation}\label{eq:XS_ION}
    \begin{split}
        \sigma_{\rm ion} =\, & \frac{4\pi\alpha^6 Z^5}{m_e \varepsilon'_{\gamma}}\left(\gamma_e^2 - 1\right)^{3/2} \Bigg[\frac{4}{3} +  \\
        &  \frac{\gamma_e(\gamma_e-2)}{\gamma_e+1}\left(1-\frac{1}{2\gamma_e\sqrt{\gamma_e^2 - 1}}\ln \frac{\gamma_e+\sqrt{\gamma_e^2-1}}{\gamma_e-\sqrt{\gamma_e^2-1}}\right) \Bigg] \,,
    \end{split}
\end{equation}
where $\gamma_e = 1+\varepsilon'_\gamma/m_e$ is the electron Lorentz factor.

Two remarks are in order: First, the PPC and photo-ionization cross sections for higher atomic shells are notably smaller than those of the $K$-shell. Specifically, the photo-ionization cross sections for the $2s$ and $2p$ sub-shells of the $L$-shell are suppressed by factors of $2^3(I_L/I_K)^3$ and $2^3(I_L/I_K)^3(I_L/\varepsilon'_\gamma)$, respectively~\cite{Akhiezer:1986yqm}. This suppression plays a pivotal role in the PPC-ionization equilibrium stage, which will be further discussed in the next section. Second, while the cross sections provided in Eqs.~\eqref{eq:XS_PPC} and \eqref{eq:XS_ION} are not viable for high charges ($\alpha Z \sim 1$), the suppression for the largest $Z$ values considered in this work is at most a factor of $\sim$ few compared to the above scaling. Thus, our findings remain robust even in absence of detailed shell-specific cross sections, whose precise calculation is beyond the scope of this exploratory study, and for which we are not aware of well-established fits or numerical tables in the literature.

Using the energy relation in Eq.~\eqref{eq:E+CoM} and CoM energy squared in the rest frame of the nuclei, $s = m_A^2 + 2\varepsilon'_\gamma m_A$, these cross sections can be expressed in terms of the Lorentz invariant variable $s$, which can be used in Eqs.~\eqref{eq:int_rate} and \eqref{eq:loss_rate}. Note that while the cross sections in Eqs.~\eqref{eq:XS_PPC} and~\eqref{eq:XS_ION} depend solely on the nuclear charge in the nucleus rest frame, independent of the atomic number or atomic mass number, the same is not true in the Lab frame: Due to the relativistic boost, the cross sections in the Lab frame depend on the atomic mass number of nucleus.

%%%%%%%%%%%%%%%%%%%%%%%%%%%%%%%%%%
%%%%%%%%%%%%%%%%%%%%%%%%%%%%%%%%%%
\subsection{Nuclei Energy Loss Processes}\label{subsec:BH_PD}
%%%%%%%%%%%%%%%%%%%%%%%%%%%%%%%%%%
%%%%%%%%%%%%%%%%%%%%%%%%%%%%%%%%%%

As nuclei propagate through a photon field, they can lose energy through various interactions such as Bethe-Heitler pair production, photo-disintegration and photo-meson production. Additionally, energy loss can occur via synchrotron radiation, due to the presence of magnetic fields in the medium. The competition between synchrotron radiation and other loss processes is influenced by the relative energy density in the magnetic field with respect to the energy density in the background photon field. Since this ratio is rather model-dependent and since the synchrotron radiation is inversely proportional to the nucleus mass, for simplicity we will ignore synchrotron losses in the following. 

For sufficiently broad photon spectra, an unavoidable and nuanced competition exists between Bethe-Heitler, photo-disintegration and photo-meson losses. This complexity stems from two facts: On the one hand, the threshold energies for the these processes in the nucleus frame are $\sim 1\,$MeV for the Bethe-Heitler, a few MeV (the exact value being nucleus-dependent) for the photo-disintegration  and $\sim 100\,$MeV/nucleon for the  photo-meson interaction. Just based on this trend, one would expect that the Bethe-Heitler matters the most at low-energies, while photo-meson interactions dominate at the highest energies, with a relevant role of the photo-disintegration in between, due to the onset of the Giant Dipole Resonance (GDR)~\footnote{For a detailed discussion on photo-nuclear interactions, see~\cite{Kossov2002pgamma, Morejon:2019pfu}.}. However, the onset energy scale for these processes depend on the nucleus since Bethe-Heitler cross section scales with $\alpha^3 Z^2$, while both the photo-disintegration and photo-meson cross sections scale approximately as $\sim A$. Furthermore, the interplay between these photon-nucleus interactions is further complicated by their inherently stochastic nature. In the context of this study, we adopt a simplified approach by defining a characteristic loss length at each energy and for each nucleus as the smallest among the Bethe-Heitler loss length, photo-meson loss length and the photo-disintegration interaction length (or timescale). Within this timescale, we will solely focus on the competition between PPC and photo-ionization processes. This approximation makes sense only if these processes are fast enough, but this is the only regime where the effects of PPC and photo-ionization cannot be neglected. Technically, we adopt the total cross section and inelasticity of the Bethe-Heitler process from~\cite{Chodorowski1992}. The interaction rate for photo-disintegration is estimated using the single-pole approximation within the Narrow-Width Approximation, as outlined in~\cite{Anchordoqui:2006pe}. For the photo-meson cross section we employ the fit in~\cite{Kossov2002pgamma}, based on GEANT4, along with the inelasticity $\eta_{\rm phmes} = 0.5$, for estimating the photo-meson loss rate.

For an ion with atomic number $Z$, mass number $A$ and charge $Z_*$ (where $Z_* \leq Z$, meaning that the nucleus is dressed by $Z-Z_*$ electrons), the PPC and ionization rates are denoted respectively by $\Gamma_{\rm PPC}(Z_*,E_A)$ and $\Gamma_{\rm ion}(Z_*,E_A)$, as functions of the charge $Z_*$ and the ion energy $E_A$. We start our treatment with a fully stripped nucleus ($Z=Z_*$); which reflects the fact that at low energies where it presumably enters the acceleration process, ionization rates exceed all the other rates of interest. Our goal is then to identify for each species, labelled by its atomic number $Z$, the energy at which the PPC rate, $\Gamma_{\rm PPC}(Z_* = Z)$, is comparable to the ionization rate when the nucleus is dressed with a single electron, $\Gamma_{\rm ion}(Z_*' = Z - 1)$, in a photon background parameterized by either a blackbody~\footnote{The results can be trivially generalized to the case where a greybody factor $\xi$ is present.} temperature $T$ or a monochromatic spectrum of energy $\epsilon_0$, occupying a region of spatial extent $R$. Only if, simultaneously, these rates surpass those of the relevant loss processes, a significant fraction of dressed nuclei then exists within the specified parameter space.

\begin{figure}[H]
\centering
\subfloat[]{
\includegraphics[width=0.5\textwidth]{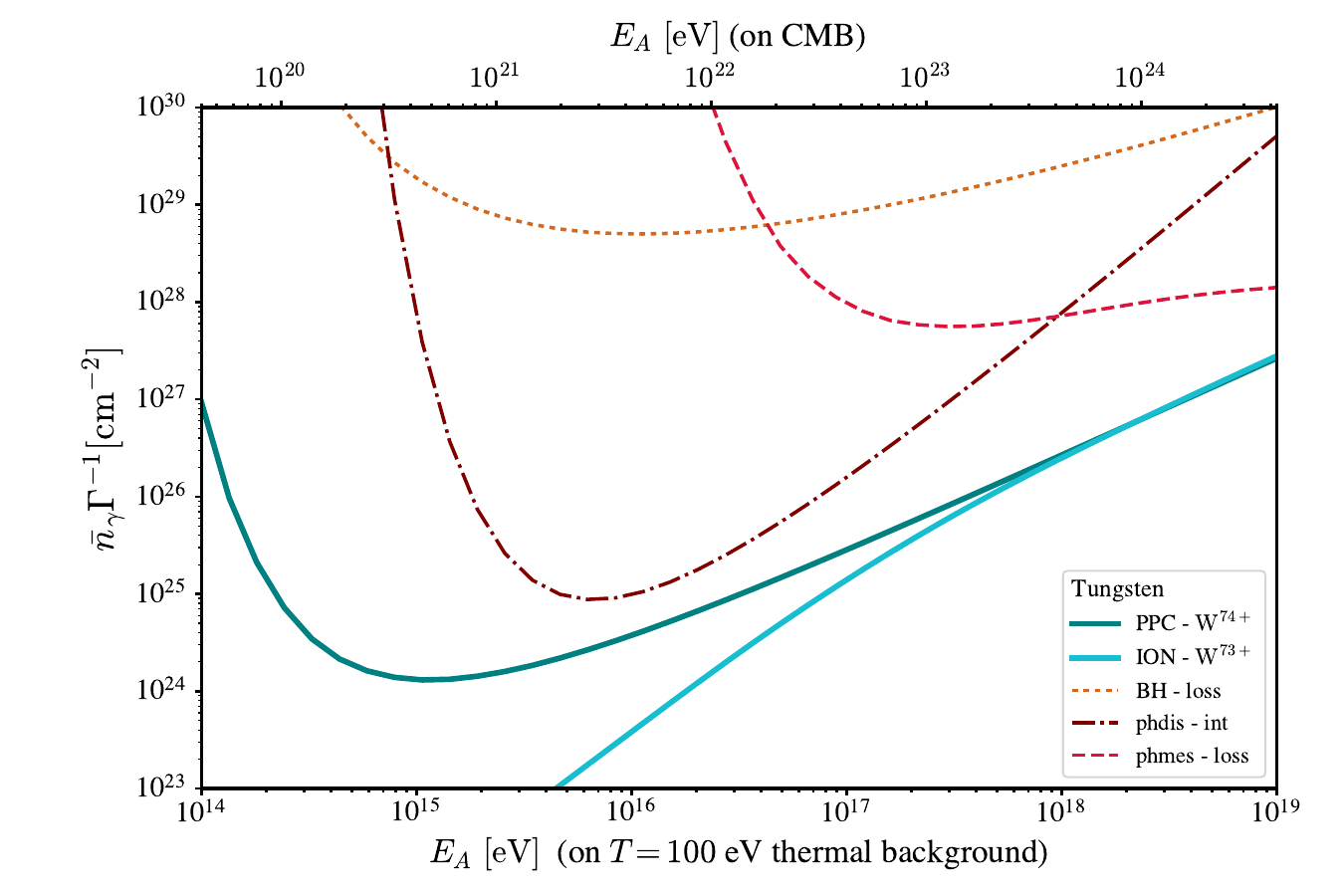}
\label{fig:len_W74}
}

\subfloat[]{
\includegraphics[width=0.5\textwidth]{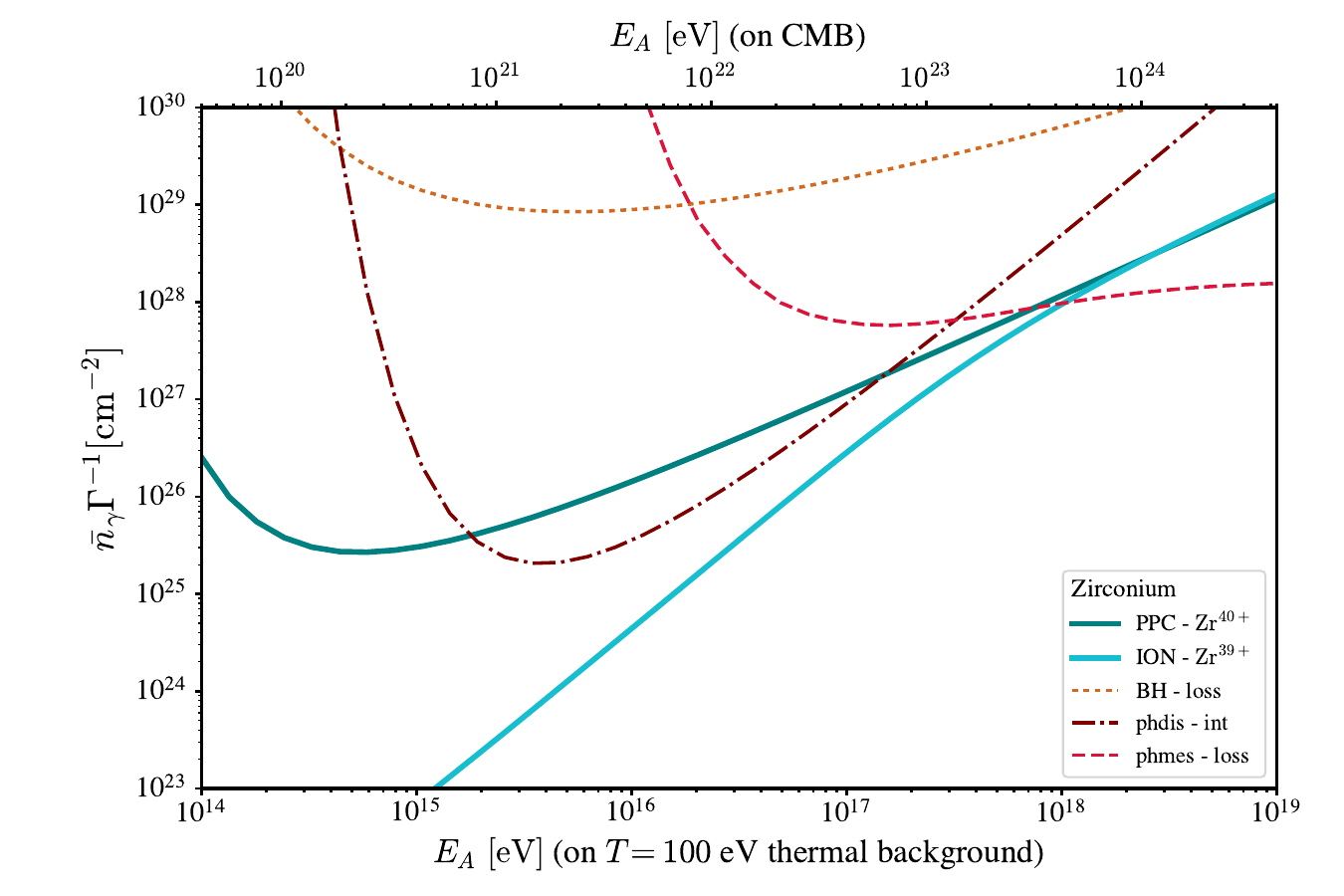}
\label{fig:len_Zr40}
}

\subfloat[]{
\includegraphics[width=0.5\textwidth]{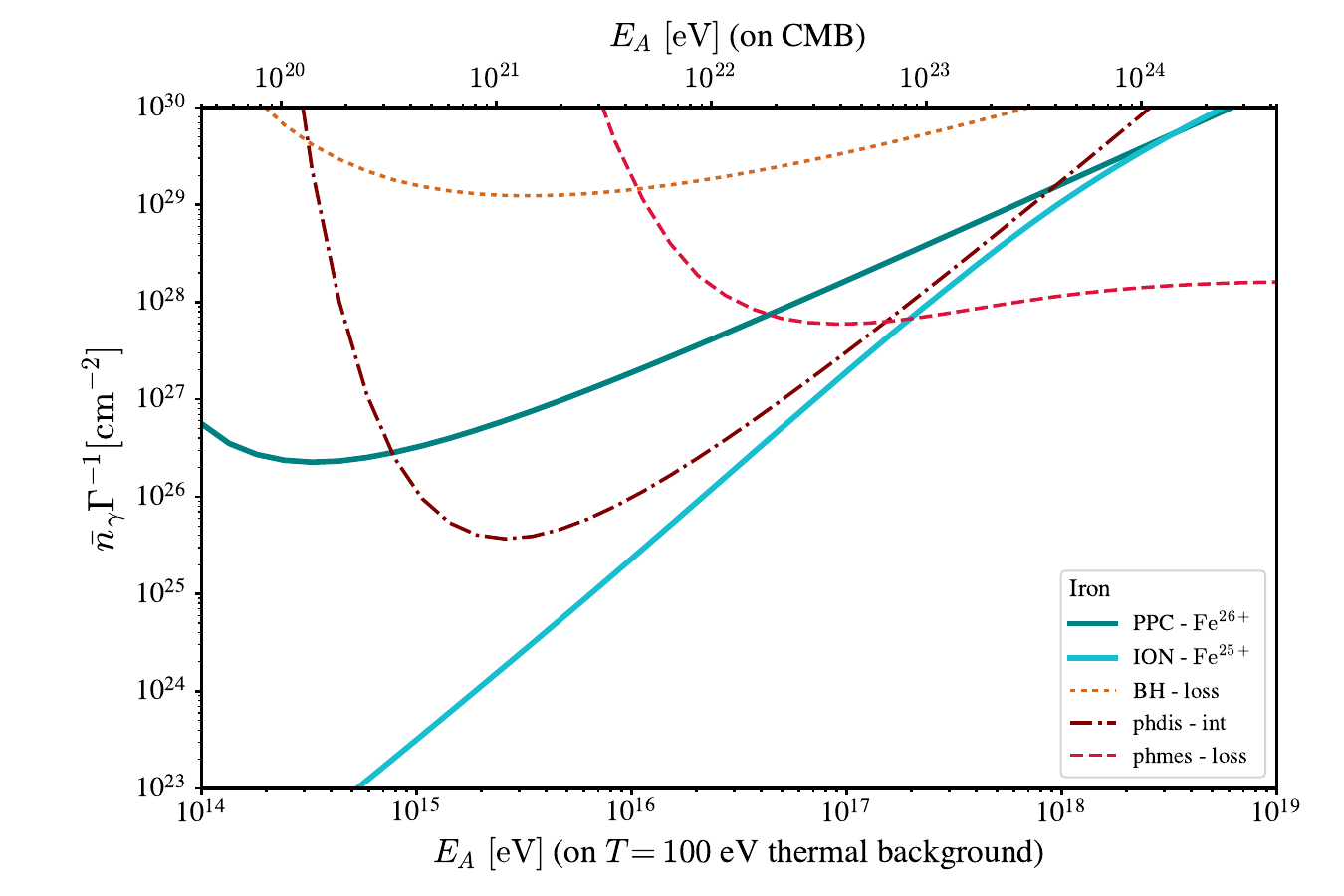}
\label{fig:len_Fe26}
}

\caption{Characteristic lengths for (a) Tungsten, (b) Zirconium and (c) Iron. The lower $x$-axis represents the nuclei's energy for the propagation inside an astrophysical source with a thermal photon field with temperature 100~eV. The upper $x$-axis corresponds to cosmological propagation in CMB. The $y$-axis shows interaction rates, normalized by the total number density of the photons, resulting in universal curves at ultra-high energies. The universality holds in the regime where $s \approx 4\varepsilon_\gamma E_A$, where $\varepsilon_\gamma$ is the photon's energy, which is valid at sufficiently high energies.}
\label{fig:lengths}
\end{figure}

The characteristic lengths of energy loss processes, as well as the PPC and ionization interaction lengths, are illustrated for both cosmological propagation and propagation within astrophysical sources in Figures~\ref{fig:len_W74}, \ref{fig:len_Zr40} and \ref{fig:len_Fe26}, respectively for Tungsten, Zirconium and Iron. For the cosmological propagation we consider interactions with the CMB photons at redshift $z=0$ (upper $x$-axis), while for the astrophysical case we use a thermal photon field with the temperature 100~eV (lower $x$-axis), both plotted against the nuclei total energy $E_A$. It should be noted that the photon field's number density, $\Bar{n}_\gamma = \int n_\gamma(\varepsilon_\gamma)\d \varepsilon_\gamma$, acts as a normalization factor for the interaction/loss lengths and scales them uniformly. As a result, the normalized inverse areas $\Bar{n}_\gamma\Gamma^{-1}$ depend only on the energy of the incident ion and the shape of the target photons energy spectrum. Given that both considered photon fields have blackbody spectrum, the normalized lengths are universal at ultra-high energy regime, as shown in Figure~\ref{fig:lengths}. As a consequence, as long as the PPC and photo-ionization interaction lengths are smaller than or comparable to the source radius $R$, the number density does not impose any further constraints.

%%%%%%%%%%%%  Figure 01  %%%%%%%%%%%%%
%%%%%%%%%%%%%%%%%%%%%%%%%%%%%%%%%%%%%%
%%%%%%%%%%%%%%%%%%%%%%%%%%%%%%%%%%%%%%
%%%%%%%%%%%%%%%%%%%%%%%%%%%%%%%%%%%%%%

In Figure~\ref{fig:lengths} two features are worth noting: {\it i}) PPC and ionization lengths converge at sufficiently high energies; {\it ii}) In the energy range where they are comparable, they are sufficiently short with respect to the energy loss lengths only if the atomic number is sufficiently large.

Thus, from Figure~\ref{fig:lengths}, we expect only a mild dressing of iron nuclei, say, but potentially large deviations from ``fully stripped'' ultra-heavy nuclei, such as Zirconium (middle panel) or Tungsten (top panel). Note that, due to the typical energy of the CMB photons in the local universe, $\varepsilon_{\rm cmb}\approx 6.3\times 10^{-4}\,$eV, a comparable PPC and ionization rate is only attained at extremely high energies, $E_A\gtrsim 10^{23}$ eV, as is manifested in Figure~\ref{fig:lengths}. Thus, {\it only a tiny fraction of dressed nuclei is expected to be observed in UHECRs propagating over cosmological distances}. Apart from the case of protons, this result is at odds with the conclusion of~\cite{Tinyakov:2006jm}, which predicts, for instance, that Fe in UHECRs should have an equilibrium $Z_*=Z-1=25$. As far as we can infer, this stems from the incorrect scaling of the photo-ionization cross section with $Z$, implicitly used in~\cite{Tinyakov:2006jm}. Instead, for the propagation inside the astrophysical sources, if the acceleration site is surrounded by a sufficiently extended environment dominated by a (not too broad) photon background, we expect that a significant fraction of partially dressed nuclei will be established.

After this qualitative assessment, in the next section we will tackle the problem more quantitatively by formulating the Saha-like equations to determine the fraction of dressed nuclei at different energies and for different elements.

%%%%%%%%%%%%%%%%%%%%%%%%%%%%%%%
%%%%%%%%%%%%%%%%%%%%%%%%%%%%%%%
\section{Saha-like Equations and Results}\label{sec:saha}
%%%%%%%%%%%%%%%%%%%%%%%%%%%%%%%
%%%%%%%%%%%%%%%%%%%%%%%%%%%%%%%

To gauge the influence of PPC and photo-ionization processes on the propagation of ultra-heavy and ultra-high-energy nuclei, we employ a set of ``Saha-like'' rate equations. These equations allow us to calculate the fraction of nuclei that remain partially dressed after traveling a given distance. The rate of change in the population of nuclei in a particular ionization state is governed by the following time-dependent set of equations:
\begin{equation}\label{eq:saha_full}
    \begin{split}
      \Dot{X}_i =\, & (1-\delta_{i0})\Gamma_{\rm ion}(i-1)X_{i-1} \\ & +  (1-\delta_{iZ})\Gamma_{\rm PPC}(i+1)X_{i+1} \\ 
      & - \left[(1-\delta_{iZ})\Gamma_{\rm ion}(i)+(1-\delta_{i0})\Gamma_{\rm PPC}(i)\right]X_i~. 
    \end{split}
\end{equation}
As before, $Z$ represents the atomic number equal to the charge of the fully stripped nucleus, while $X_i$ refers to the fraction of nuclei with same mass number $A$ in an ionization state corresponding to charge $i$, where $0 \leq i \leq Z$, and the system must be solved from top ($X_{Z}$, fully stripped) to bottom ($X_{0}$, fully dressed). 

We begin by setting the boundary condition $X_i(0) = \delta_{iZ}$, which means that initially all the nuclei are fully stripped. Additionally, we impose the constraint $\sum_i X_i = 1$, ensuring that the total population of nuclei remains conserved across different ionization states. The system of differential equations is then solved over the time interval $(0,\Gamma_{\rm max}^{-1})$, with $\Gamma_{\rm max}$ being the most efficient loss rate at the considered energy for the specific nucleus. This ensures that the conservation of the number of ions of each species is approximately valid, and also allows us to solve the equations for constant values of energy. 

Formally, Eq.~\eqref{eq:saha_full} admits a steady state solution if all the time derivatives vanish after a {\it relaxation} time scale $\tau_{\rm rel}$, {\it i.e.} $\Dot{X}_i(t \gtrsim \tau_{\rm rel}) = 0$. In the steady state the population of nuclei in each ionization state remains constant over time and the the solution of the system of equations in Eq.~\eqref{eq:saha_full}, for $t \gtrsim \tau_{\rm rel}$, takes the following form:
\begin{equation}\label{eq:eq:saha_steady_full}
    X_{Z-n} = \prod_{i=0}^{n-1}\frac{\Gamma_{\rm PPC}(Z-i)}{\Gamma_{\rm ion}(Z-i-1)}\,.
\end{equation}
Clearly, the steady state can be achieved if {\it all} the PPC and photo-ionization rates exceed the loss rates, allowing the nuclei to reach a stable ionization state. The relaxation time can then be estimated by $\tau_{\rm rel}\sim \left[\sum_{i=0}^{Z-1}\Gamma^{-1}_{\rm PPC}(Z-i) + \Gamma^{-1}_{\rm ion}(Z-i-1)\right]$. It is evident that the set of equations in Eq.~\eqref{eq:saha_full} may never have enough time to reach the steady state in a physical situation, because of the significantly suppressed PPC and photo-ionization rates in the higher atomic shells, as previously discussed. In the absence of the high-shell suppression, the PPC and ionization rates would be comparable and the system would quickly achieve a steady state with a ``democratic'' distribution of dressed nuclei, that is $X_i\sim1/Z$ for all $i$.

In practice and for the physical conditions we are dealing with in this work, we can safely assume that $X_{i<Z-4}\approx0$, allowing us to truncate the system of equations in Eq.~\eqref{eq:saha_full} at $X_{Z-3}$ and simplify the last equation to
\begin{equation}
     \Dot{X}_{Z-3} = -\Gamma_{\rm ion}(Z-3)X_{Z-3}+\Gamma_{\rm PPC}(Z-2)X_{Z-2}\,. 
\end{equation}
For the purposes of this study, we rely on the $K$-shell PPC and photo-ionization cross sections. Additionally, we incorporate a suppression factor of $1/8$ for the $(Z-3)$ ionization state with respect to the $(Z-2)$ one. 

\begin{figure}[H]
\centering
\subfloat[]{
\includegraphics[width=0.5\textwidth]{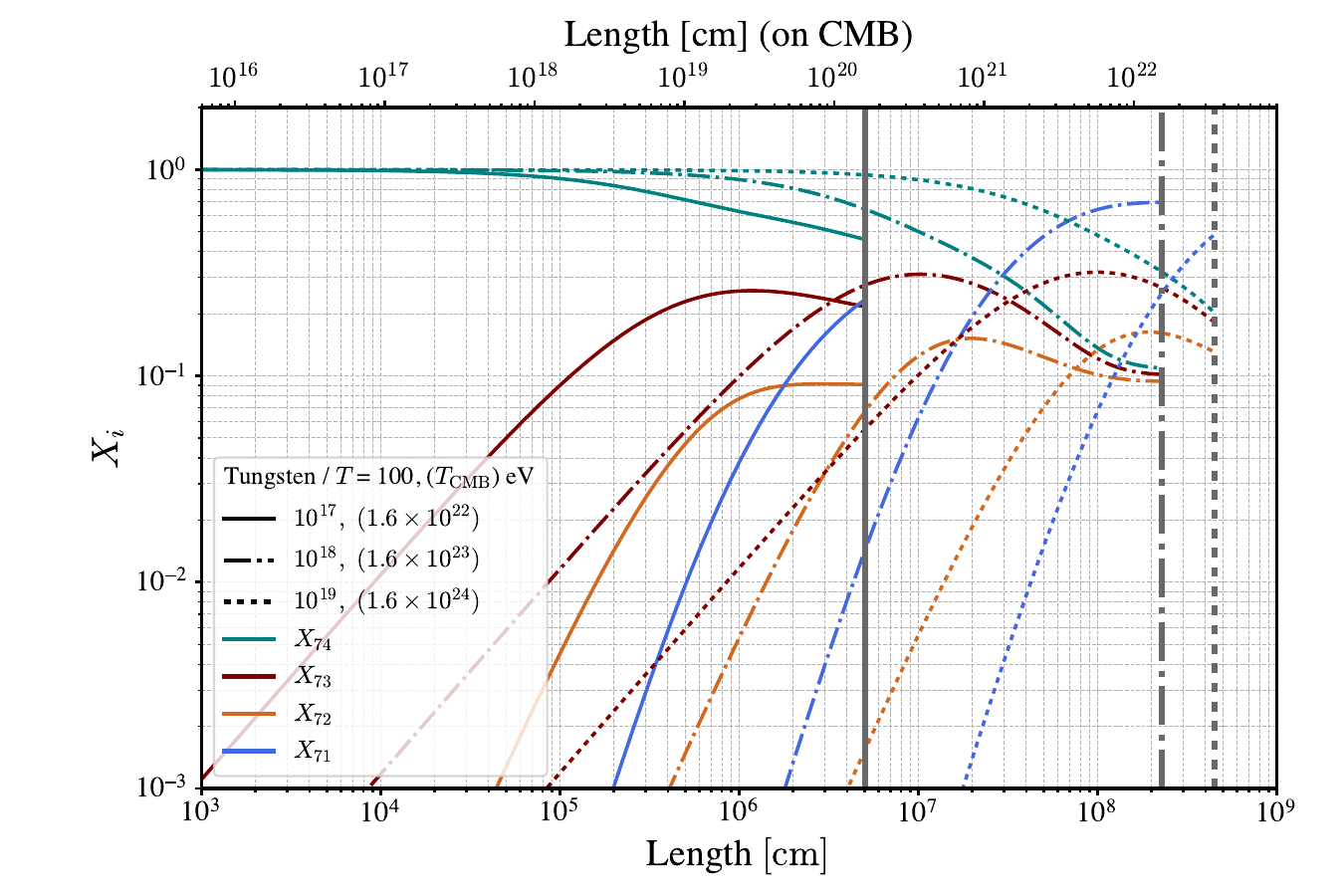}
\label{fig:Xf_W_ThX100}
}

\subfloat[]{
\includegraphics[width=0.5\textwidth]{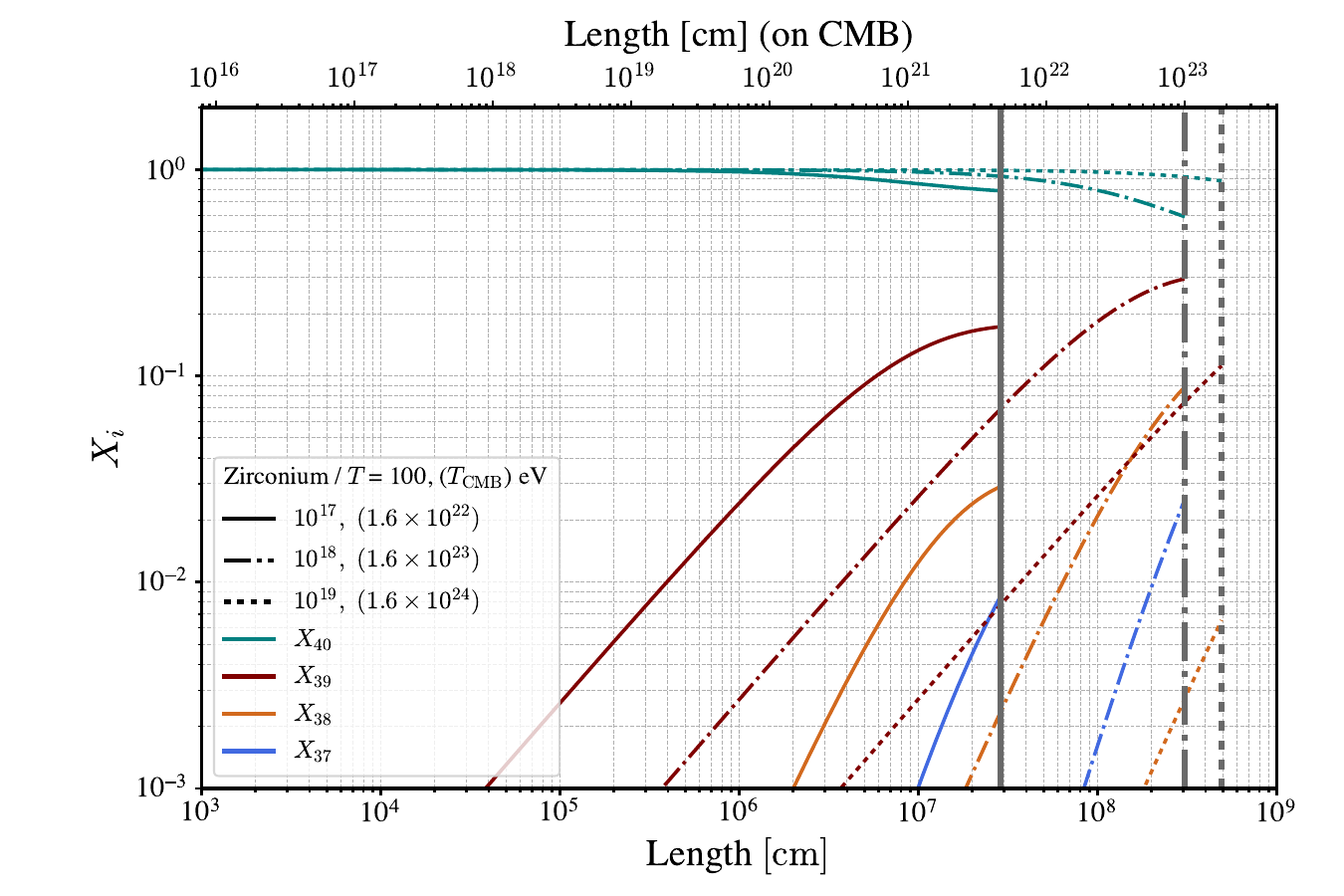}
\label{fig:Xf_Zr_ThX100}
}

\subfloat[]{
\includegraphics[width=0.5\textwidth]{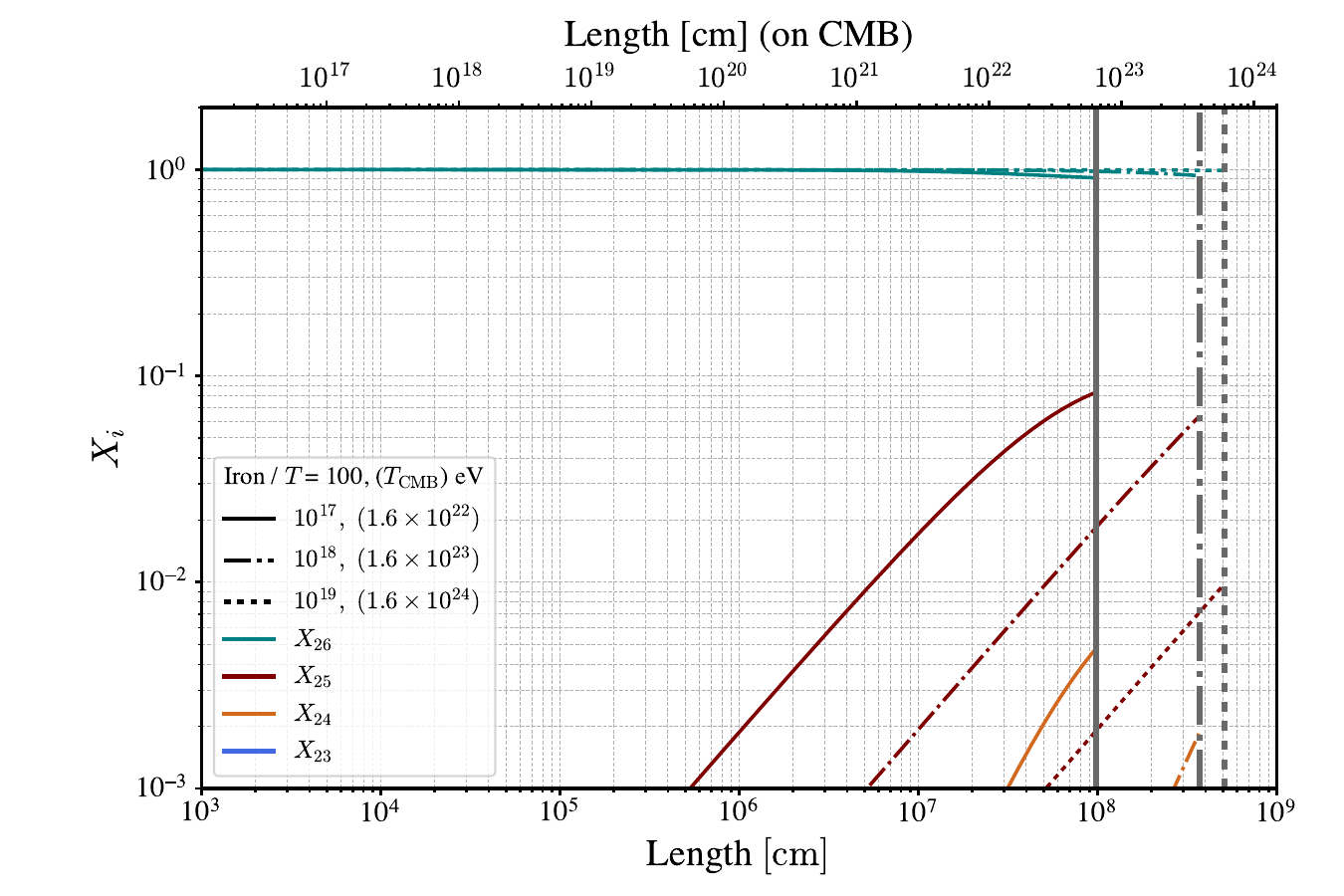}
\label{fig:Xf_Fe_ThX100}
}

\caption{Fraction of nuclei dressed with one, two and three electrons, from numerical solution of Eq.~\eqref{eq:saha_full} for (a) Tungsten, (b) Zirconium and (c) Iron, propagating in a thermal photon field of temperature $T=100$~eV, with energies $E_A = 10^{17}$, $10^{18}$ and $10^{19}$~eV. The vertical thick gray lines illustrate the minimum loss length for the corresponding energies. For propagation in CMB the dressing fractions can be achieved for nuclei with energies reported in the legends and over distances in the upper $x$-axis.}
\label{fig:Xf_ThX100}
\end{figure}

\begin{figure}[t!]
\centering
\subfloat[]{
\includegraphics[width=0.5\textwidth]{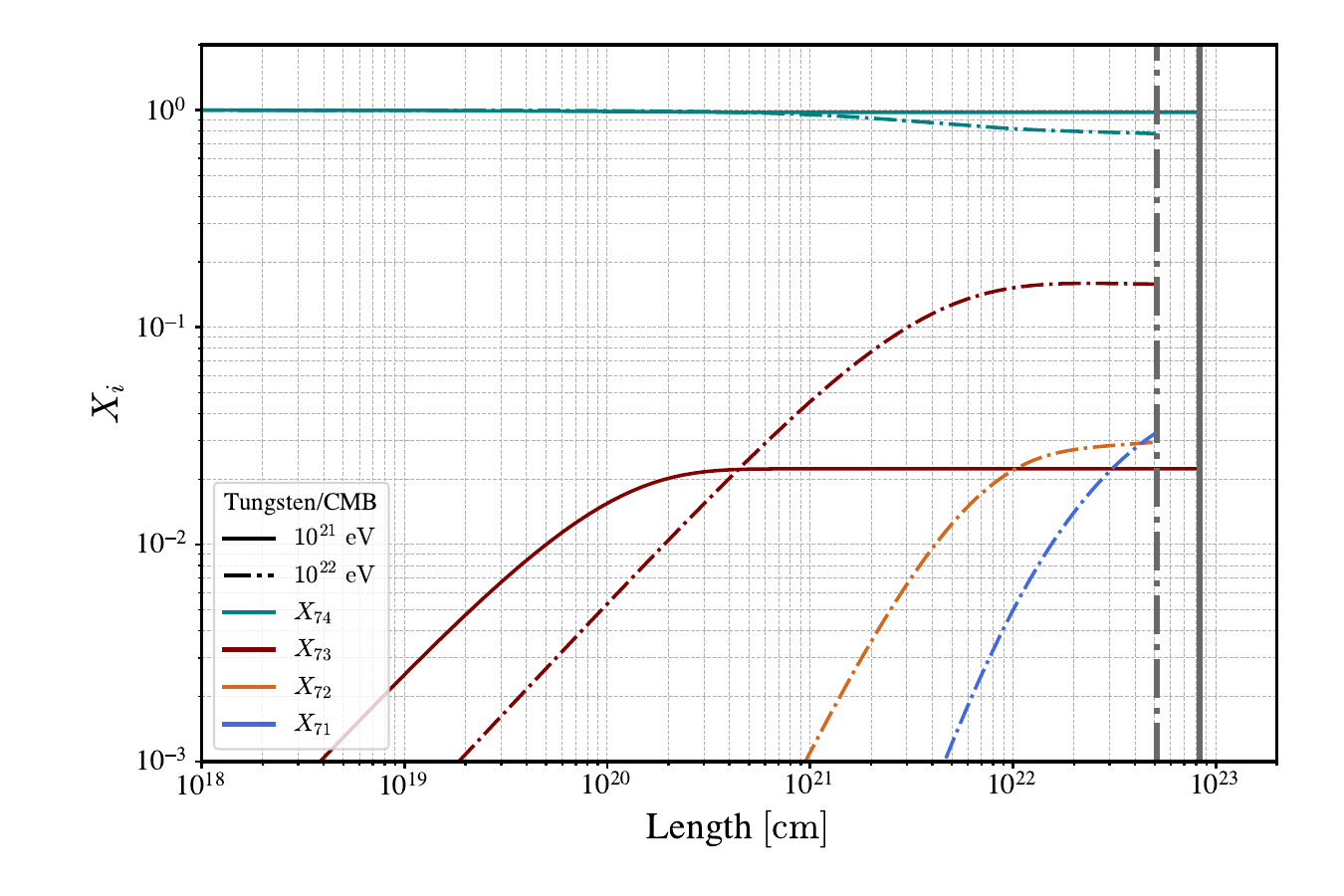}
\label{fig:Xf_W_CMB}
}

\subfloat[]{
\includegraphics[width=0.5\textwidth]{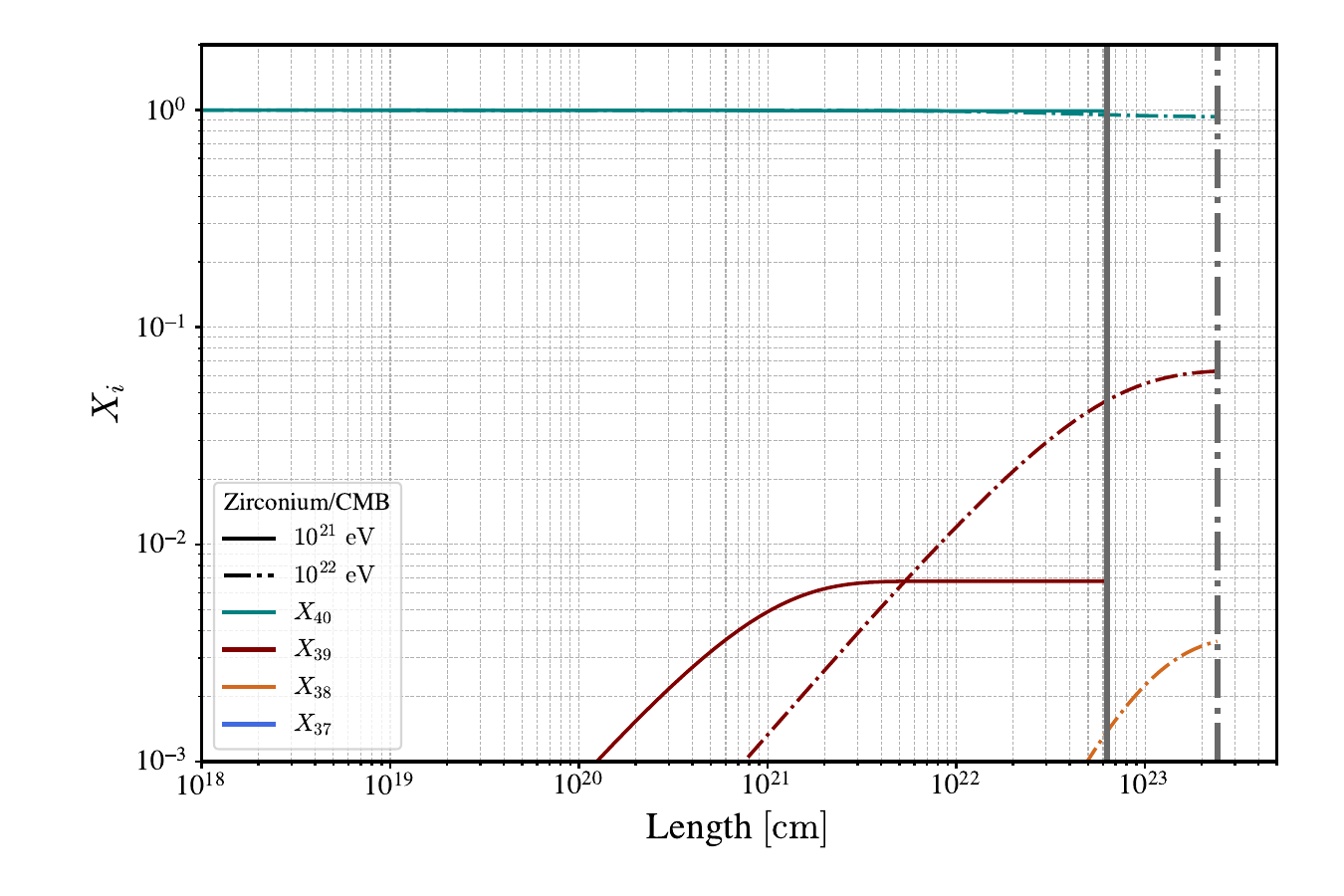}
\label{fig:Xf_Zr_CMB}
}

\caption{Same as Figure~\ref{fig:Xf_ThX100}, for (a) Tungsten and (b) Zirconium nuclei propagating in the CMB and with energies $E_A = 10^{21}$ and $10^{22}$~eV.}
\label{fig:Xf_CMB}
\end{figure}

For our truncated system of equations, the steady state solution is attainable (i.e., can be achieved for ultra-heavy nuclei, such as Tungsten), and takes the form:
\begin{equation}\label{eq:saha_steady}
  \begin{aligned}
    &X_{Z-1}=\frac{\Gamma_{\rm PPC}(Z)}{\Gamma_{\rm ion}(Z-1)}~, \\
    &X_{Z-2}=\frac{\Gamma_{\rm PPC}(Z-1)}{\Gamma_{\rm ion}(Z-2)}\frac{\Gamma_{\rm PPC}(Z)}{\Gamma_{\rm ion}(Z-1)}~, \\
    &X_{Z-3}=\frac{\Gamma_{\rm PPC}(Z-2)}{\Gamma_{\rm ion}(Z-3)}\frac{\Gamma_{\rm PPC}(Z-1)}{\Gamma_{\rm ion}(Z-2)}\frac{\Gamma_{\rm PPC}(Z)}{\Gamma_{\rm ion}(Z-1)}~. 
\end{aligned}  
\end{equation}
This solution aligns well with the numerical solution of the Saha-like equations shown in Figure~\ref{fig:Xf_ThX100} for the propagation of nuclei in a blackbody photon field of temperature 100~eV. Figure~\ref{fig:Xf_W_ThX100} shows that for a Tungsten nucleus, $Z=74$, with the energy $E_A = 10^{17}$~eV and propagating through the blackbody photon field of temperature 100~eV, the fractions of dressed nuclei with one, two or three electrons reach $\{X_{73}, X_{72}, X_{71}\}=\{0.22, 0.09, 0.23\}$ until the smallest energy loss length indicated by the vertical gray solid line, meaning that nearly half of the initially bare nuclei become dressed. By increasing the energy to $E_A = 10^{18}$~eV ($10^{19}$~eV) the fraction of dressed nuclei, $1-X_{74}$, increases to approximately $89\%$ ($79\%$). For Zirconium in Figure~\ref{fig:Xf_Zr_ThX100}, the effect is smaller but is not negligible and $1-X_{40}$ reaches $\sim 40\%$ at $10^{18}$~eV. Going to Iron in Figure~\ref{fig:Xf_Fe_ThX100}, the effect is less pronounced but still $1-X_{26}\simeq X_{25}\sim 10\%$. The upper $x$-axis in the panels of Figure~\ref{fig:Xf_ThX100} shows the required propagation length in the CMB for achieving the same dressing fractions at the energies displayed in the legends. For example, $89\%$ of Tungsten nuclei dress in the propagation through CMB at the energy $1.6\times10^{23}$~eV, which is extremely large by the Hillas criterion. The dressing fractions that can be achieved for nuclei propagating in CMB with highest observed energies are shown in Figure~\ref{fig:Xf_CMB}. For example, from Figure~\ref{fig:Xf_W_CMB}, a Tungsten nucleus escaping its source fully stripped and with $E_A = 10^{22}$~eV would arrive at Earth with $\sim 16\%$ probability of being dressed once and $\sim 3\%$ of acquiring two electrons. The Iron is not shown in Figure~\ref{fig:Xf_CMB} since the dressing probability is negligible; only $\sim 0.3\%$ of Iron nuclei at $E_A = 10^{21}$~eV dress with one electron in the course of propagation in CMB. As anticipated, we estimate that the dressing fraction associated to the extragalactic propagation is thus negligible over the range of UHECR energies observed.

The maximum dressing fraction, ${1-X_{Z}}$, for propagation in a blackbody photon field of temperature $T$ and the energy of nucleus at which it occurs for elements heavier than Iron are reported in Table~\ref{tab:max_dress}. For this case, the CoM energy squared at ultra-high energies is approximately $s \approx E_A T$. Thus, by choosing a temperature, the $E_A$ values can found from the reported value of $E_AT$ in the table.

\begin{figure}[t!]
\centering
\subfloat[]{
\includegraphics[width=0.5\textwidth]{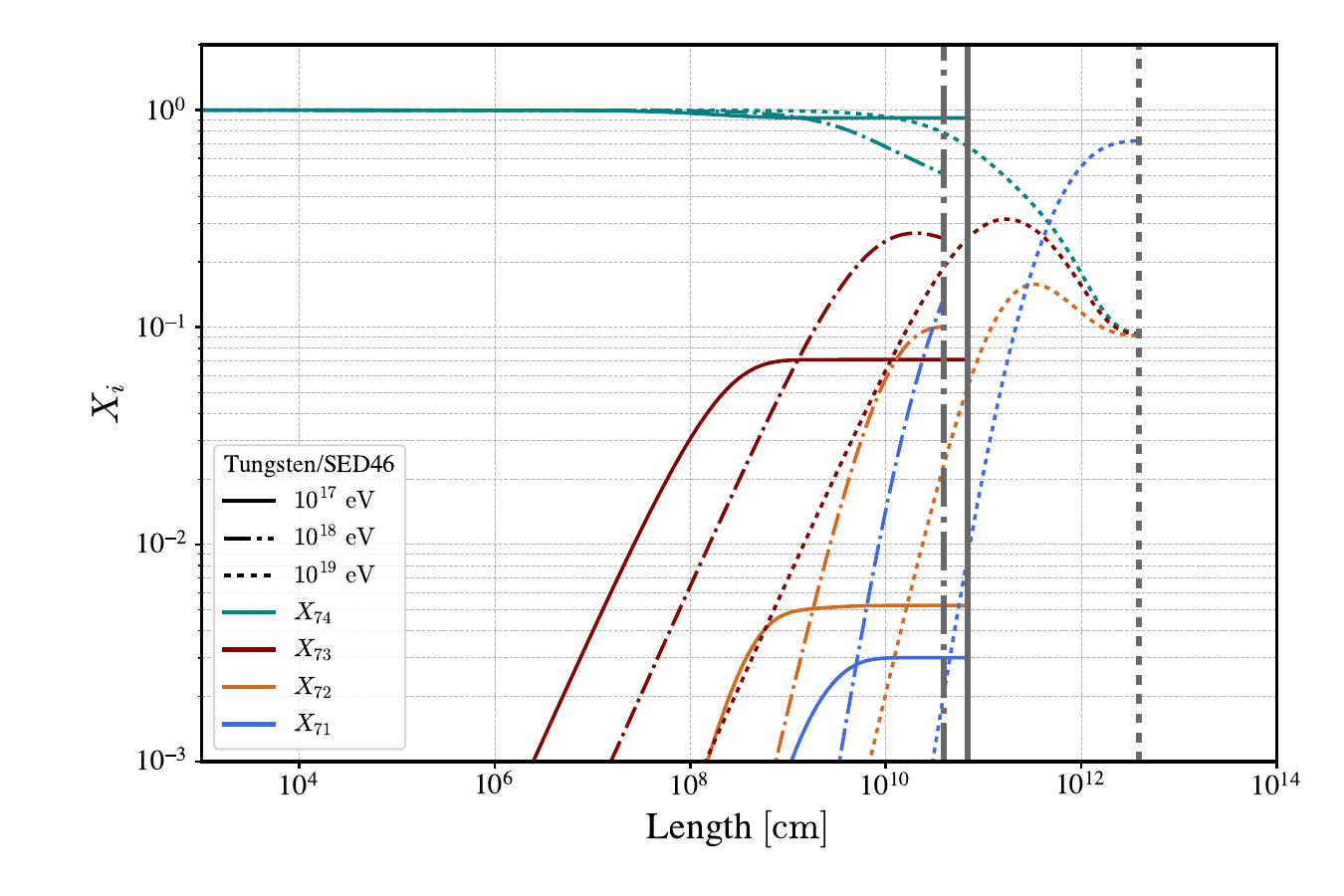}
\label{fig:Xf_W_SED46}
}

\subfloat[]{
\includegraphics[width=0.5\textwidth]{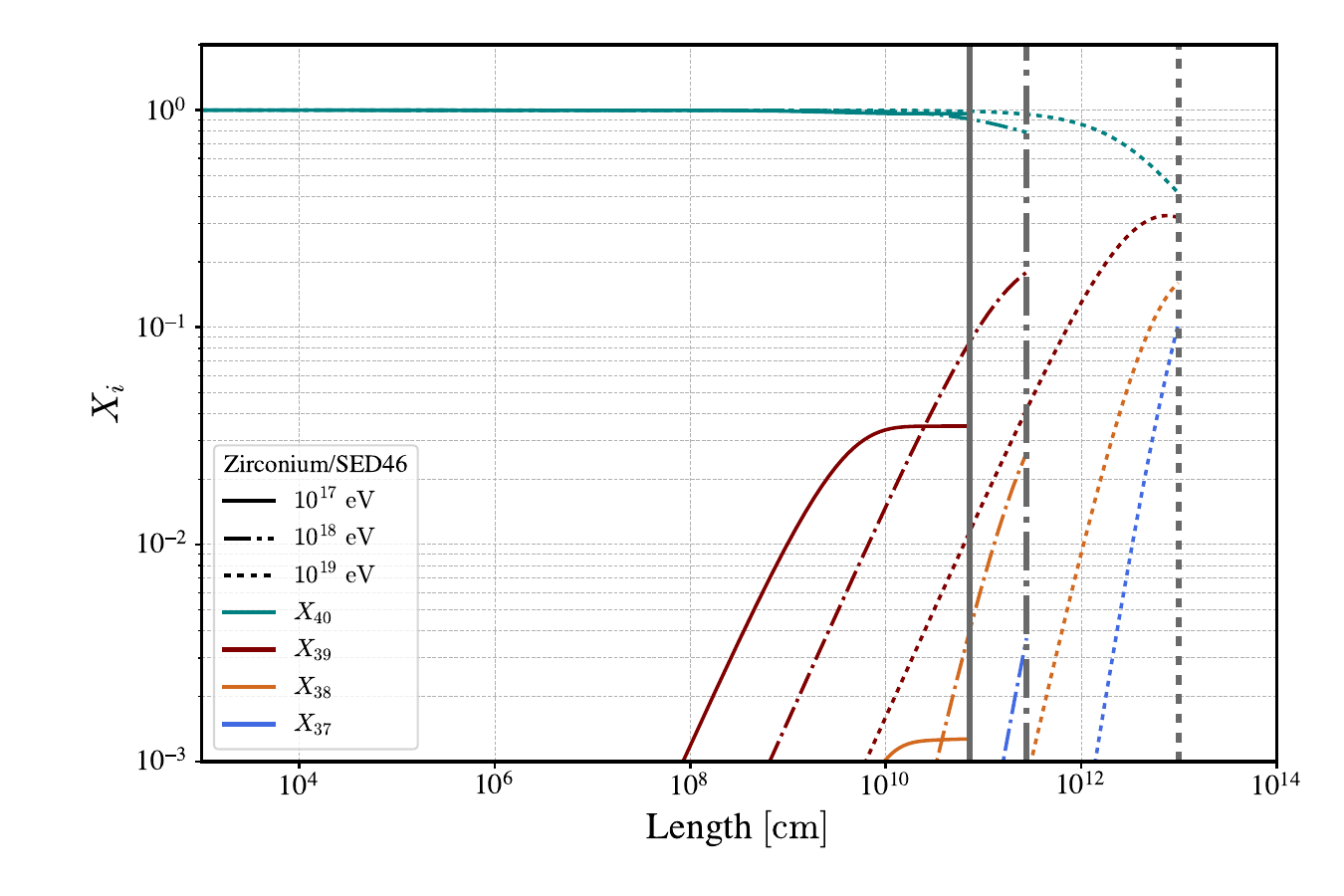}
\label{fig:Xf_Zr_SED46}
}

\subfloat[]{
\includegraphics[width=0.5\textwidth]{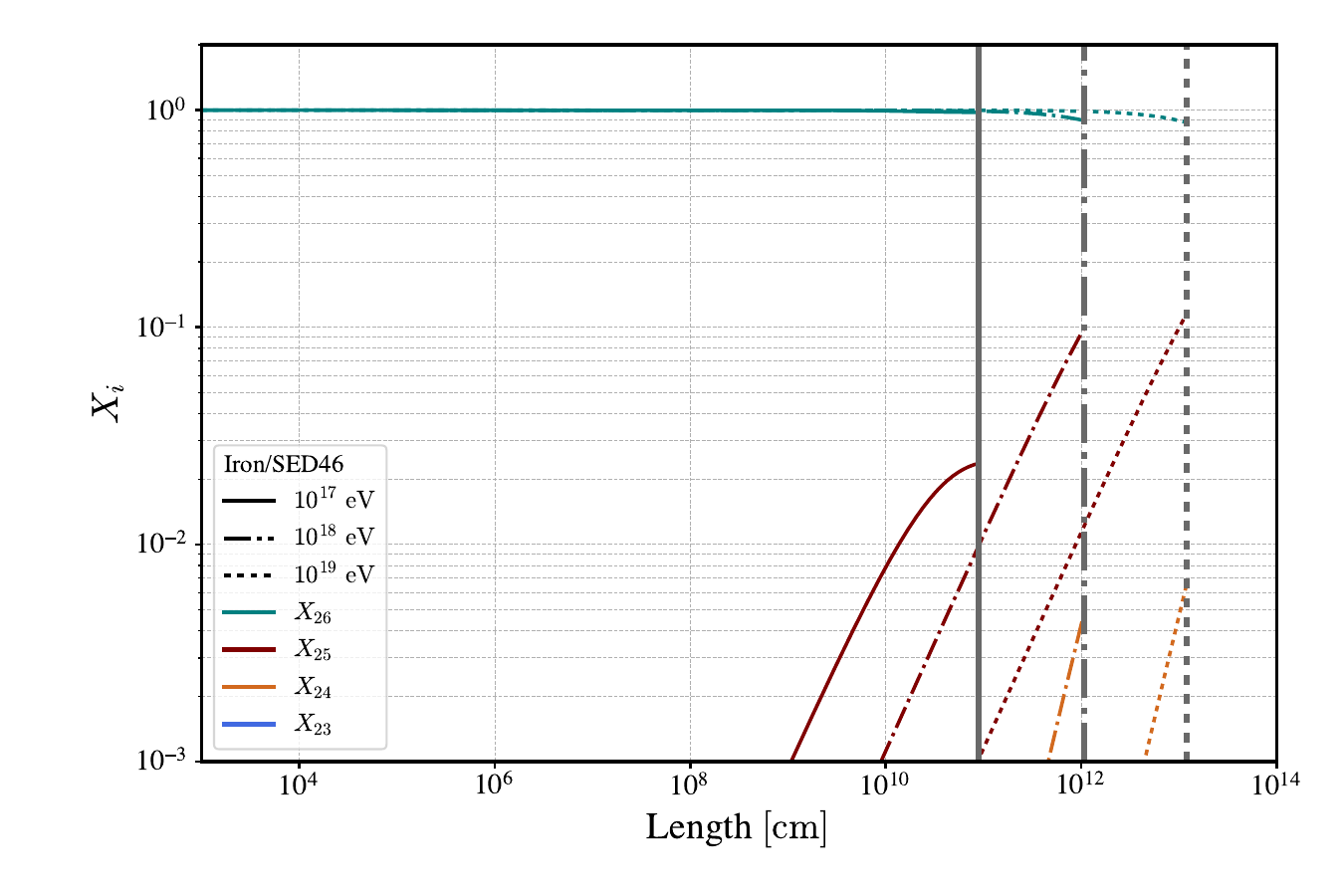}
\label{fig:Xf_Fe_SED46}
}

\caption{Same as Figure~\ref{fig:Xf_ThX100}, but for propagation within a source characterized by an SED with X-ray luminosity $L_X = 10^{46}$~erg/s, taken from~\cite{Murase:2019vdl}.}
\label{fig:Xf_SED46}
\end{figure}

Can the results obtained assuming a blackbody photon spectrum be considered representative? To answer, we turn to a state-of-the-art Active Galactic Nuclei (AGN) disk-corona scenario~\cite{Murase:2019vdl} and use the spectral energy distribution (SED) with an X-ray luminosity of $L_X = 10^{46}~{\rm erg/s}$, dubbed SED46. The SED46 has a power-law spectrum for the photon field, marking a significant departure from the blackbody spectrum examined earlier (see~\cite{Murase:2019vdl} for further details on SED46). Figure~\ref{fig:Xf_SED46} shows the fraction of dressed nuclei, same as Figure~\ref{fig:Xf_ThX100} but for propagation in the SED46 photon field. Comparison between Figures~\ref{fig:Xf_SED46} and \ref{fig:Xf_ThX100} shows an almost equal dressing fractions, pointing to the robustness of results with respect to the assumption about the photon field spectrum. 
%%%%%%%%%%%%  Figure 02  %%%%%%%%%%%%%
%%%%%%%%%%%%%%%%%%%%%%%%%%%%%%%%%%%%%%
%%%%%%%%%%%%%%%%%%%%%%%%%%%%%%%%%%%%%%
%%%%%%%%%%%%%%%%%%%%%%%%%%%%%%%%%%%%%%

%%%%%%%%%%%%  Figure 03  %%%%%%%%%%%%%
%%%%%%%%%%%%%%%%%%%%%%%%%%%%%%%%%%%%%%
%%%%%%%%%%%%%%%%%%%%%%%%%%%%%%%%%%%%%%
%%%%%%%%%%%%%%%%%%%%%%%%%%%%%%%%%%%%%%

\begin{figure}[t!]
\centering
\subfloat[]{
\includegraphics[width=0.5\textwidth]{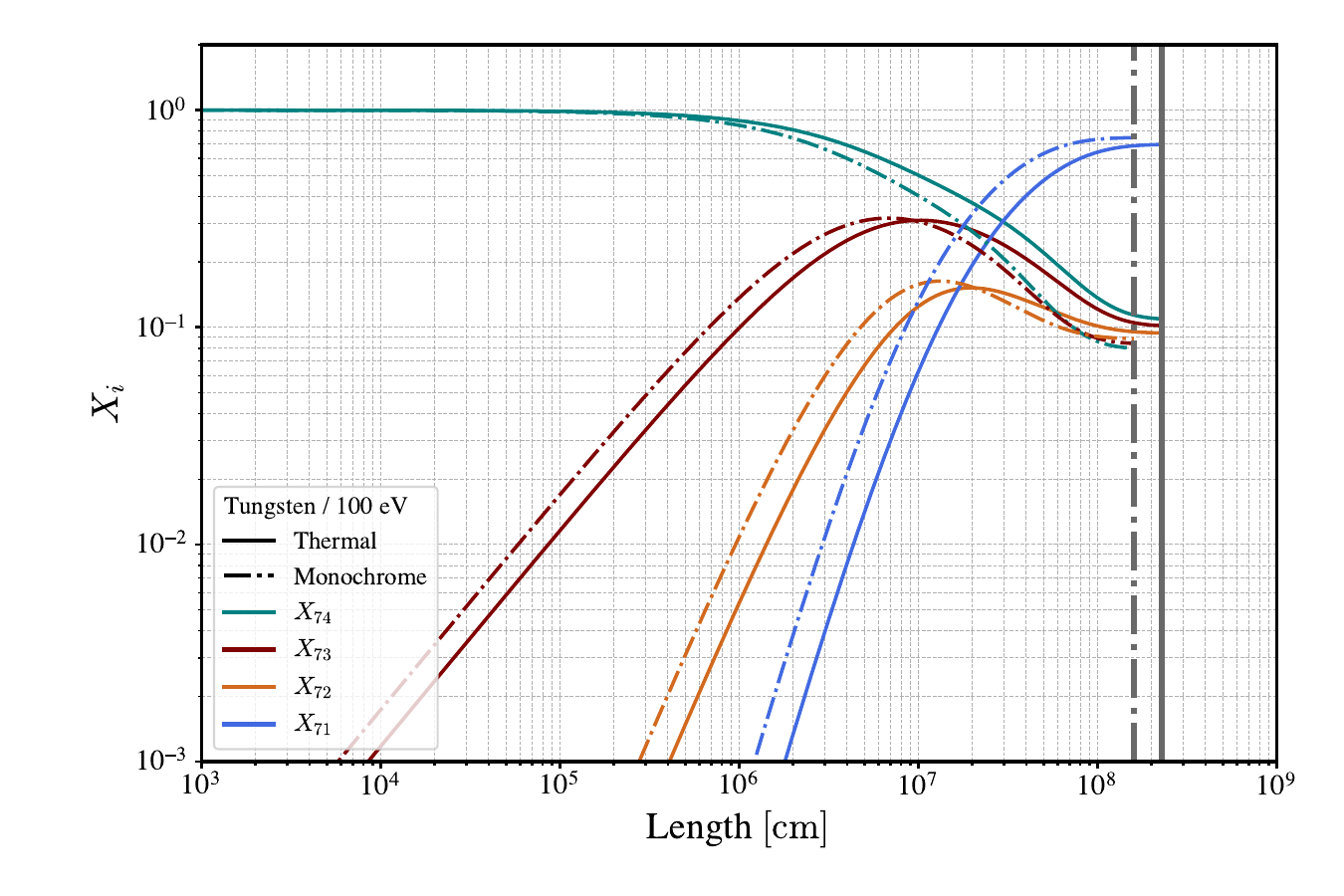}
\label{fig:Xf_W_BBmono}
}

\subfloat[]{
\includegraphics[width=0.5\textwidth]{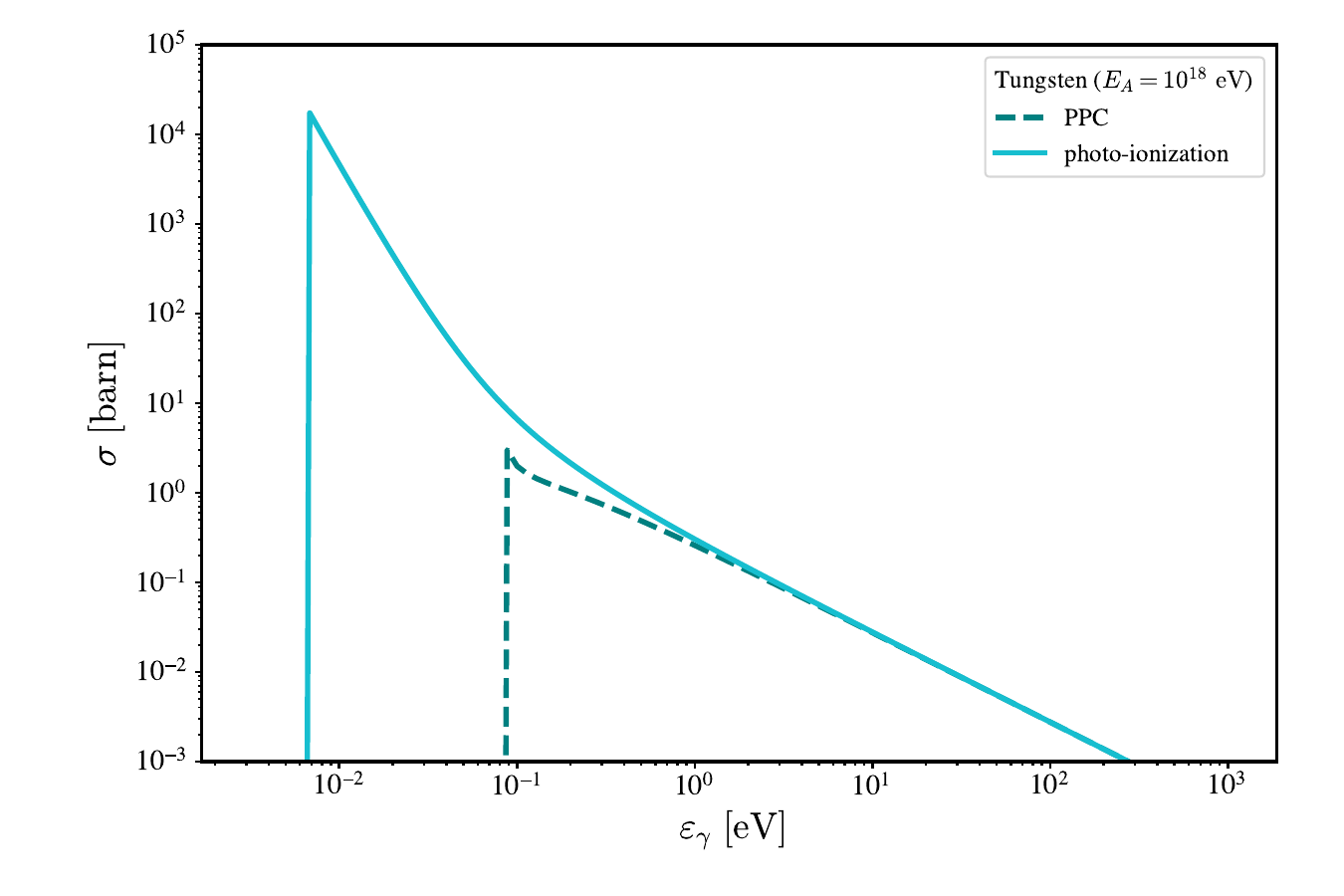}
\label{fig:xsec-PPCion}
}

\caption{\label{fig:Xf_BBmonotot}Panel (a): The fraction of dressed Tungsten nuclei, comparing the case of propagation in a blackbody spectrum of temperature $T=100$~eV with a monochromatic photon field of energy 100~eV. The vertical lines correspond to $E_A=10^{18}$~eV. The number density is fixed. Panel (b): The cross sections of PPC and photo-ionization processes as function of the photon energy in the Lab frame.}
\end{figure}

A close inspection reveals that the most important feature that may alter the obtained results is the broadness of the background photon distribution. To illustrate this, in Figure~\ref{fig:Xf_W_BBmono} we compare the dressing fractions for Tungsten propagating in blackbody of $T=100$~eV and monochromatic $\varepsilon_\gamma=100$~eV photon fields. The vertical solid and dot-dashed lines correspond to $E_A=10^{18}$~eV. An increase of a few percent in the fraction of dressed nuclei can be noticed for propagation in monochromatic photon field. Qualitatively, the reason for this increase is that the photons in the low-energy tail of the blackbody spectrum contribute more to the photo-ionization than the PPC process and therefore suppress the fraction of dressed nuclei. This can be understood by inspecting Figure~\ref{fig:xsec-PPCion}, which shows the dependence of PPC and photo-ionization cross sections for Tungsten on photon energy $\varepsilon_\gamma$ in the lab frame, with the former being penalized by a higher energy threshold and, initially, a milder decrease with energy: Due to the photo-ionization cross section growth at low-energy, even relatively few photons can move the balance in favor of the ionization.

%%%%%%%%%%%%  Figure 04  %%%%%%%%%%%%%
%%%%%%%%%%%%%%%%%%%%%%%%%%%%%%%%%%%%%%
%%%%%%%%%%%%%%%%%%%%%%%%%%%%%%%%%%%%%%
%%%%%%%%%%%%%%%%%%%%%%%%%%%%%%%%%%%%%%

%%%%%%%%%%%%  Figure 05  %%%%%%%%%%%%%
%%%%%%%%%%%%%%%%%%%%%%%%%%%%%%%%%%%%%%
%%%%%%%%%%%%%%%%%%%%%%%%%%%%%%%%%%%%%%
%%%%%%%%%%%%%%%%%%%%%%%%%%%%%%%%%%%%%%

We close this section by a brief discussion about the effect of PPC on the energy loss rate of nuclei. The PPC process adds to the Bethe-Heitler process, and thus modifies the overall energy loss incurred by heavy nuclei. This modification manifests in two distinct ways, termed here as direct and indirect ways.

The direct modification to energy loss rate can be estimated by assuming a similar inelasticity for both PPC and Bethe-Heitler processes. Then, the loss rate due to PPC can be thought of as a correction to the Bethe-Heitler loss rate given by $\sigma_{\rm PPC}/\sigma_{\rm BH} \sim [9\pi(\alpha Z)^3 m_e/(7 \varepsilon^\prime_\gamma \ln(2\varepsilon^\prime_\gamma/m_e))]$. At high energies the photo-disintegration dominates the energy loss rate for heavy nuclei except for a window at $E_A\sim\,{\rm few}\,\times10^{20}$~eV (see figure 5 in~\cite{Zhang:2024sjp}) where Bethe-Heitler energy loss length is marginally smaller, for interaction with CMB. This corresponds to $\varepsilon^\prime_\gamma\sim10m_e$, which from our estimate leads to $\sim2\%$ ($0.3\%$) modification of Bethe-Heitler loss rate for Tungsten (Zirconium).

Indirectly, PPC alters the loss rate of nuclei by modifying the ion’s charge state, which the Bethe-Heitler energy loss depends on it quadratically. For the fraction $X_{Z-1}$ of nuclei that acquire one bound electron, the Bethe-Heitler energy loss is reduced by a factor of $(1 - 1/Z)^2$. For ultra-heavy nuclei, this suppression is minimal, of the order of $\lesssim 10^{-3}$, and is therefore negligible in most cases.  

%%%%%%%%%%%%%%%%%%%%%%%%%%%%%%%%
%%%%%%%%%%%%%%%%%%%%%%%%%%%%%%%%
\section{Conclusion and Discussion}\label{sec:conclusion}
%%%%%%%%%%%%%%%%%%%%%%%%%%%%%%%%
%%%%%%%%%%%%%%%%%%%%%%%%%%%%%%%%

In the ongoing efforts to understand the origin and acceleration mechanisms of UHECRs, this study revisits a crucial but often overlooked aspect, the ionization state of nuclei, both during acceleration inside astrophysical sources and during cosmological propagation. We demonstrated that processes such as pair production with capture (PPC) and photo-ionization, while negligible for light nuclei ($Z\lesssim 26$), have a significant impact on the propagation of heavy and ultra-heavy nuclei at ultra-high energies in the presence of background photon fields. We find that a significant fraction of these nuclei become dressed during their journey through dense photon fields, accumulating electrons in the innermost shells, changing their effective electric charge. A summary of the dressing fractions for different nuclei is reported in Table~\ref{tab:max_dress}. 

The reduction in the electric charge during acceleration also reduces the overall efficiency of the acceleration process for heavy and ultra-heavy nuclei. While our analysis does not incorporate acceleration rates, it is important to note that the effect of PPC and ionization becomes more pronounced at higher energies. As the energy of the nuclei increases, so too does the probability of these interactions. Accounting for acceleration rates would likely increase the dressing fraction and its influence on UHECR propagation and energy losses.

Our results further indicate that cosmological propagation of UHECRs through the CMB does not significantly affect the ionization state of these nuclei. For the energy ranges of interest in UHECRs, photo-ionization remains the dominant process and we expect most UHECRs to arrive at Earth in a fully stripped state. Even if nuclei are emitted in a partially dressed state from their source, they will quickly become fully ionized due to the efficiency of photo-ionization in their extragalactic propagation. However, should nuclei be accelerated to extremely high energies, that is $E_A \gtrsim 10^{22}$~eV, the impact of PPC and photo-ionization would become significant even in cosmological propagation, yielding ${\mathcal O}$(10\%) of the population to be dressed. Additionally, we have shown that the PPC process contributes to a little extent to Bethe-Heitler energy losses for UHECRs. While a secondary effect, it can reach the few percent level for cosmic rays with energies of the order of a few $\times 10^{20}$~eV propagating through the CMB. This motivates the inclusion of PPC losses in precision modeling of UHECR energy loss mechanisms.

Future work will focus on a more precise assessment of the ionization states of heavy nuclei, which requires detailed calculations of PPC and photo-ionization cross sections for different atomic shells and charge states. Having these cross sections, the problem can be addressed using Monte Carlo simulations or semi-analytical methods. One interesting scenario involves nuclei capturing electrons directly into the $L$-shell without first occupying the $K$-shell. In such cases, the de-excitation of the captured electron could produce a gamma-ray signal characteristic of heavy UHECRs, which represents another promising direction for future UHECR searches, analogous to what discussed in~\cite{Murase:2010va, Kusenko:2011tb}.

%%%%%%%%%%%%%%%%%%%%%%%%%%%
\begin{acknowledgments}
%%%%%%%%%%%%%%%%%%%%%%%%%%%
The authors would like to thank Kohta Murase for feedback on an ealier version of this article. A.F.~E. acknowledges support by the Fundação Carlos Chagas Filho de Amparo à Pesquisa do Estado do Rio de Janeiro (FAPERJ) scholarship No. 201.293/2023, the Conselho Nacional de Desenvolvimento Científico e Tecnológico (CNPq) scholarship No. 140315/2022-5 and by the Coordenação de Aperfeiçoamento de Pessoal de Nível Superior (CAPES)/Programa de Excelência Acadêmica (PROEX) scholarship No. 88887.617120/2021-00. 

\end{acknowledgments}

\bibliography{refs}

%%%%%%%%%  TABLE 1  %%%%%%%%%%%%%%
%%%%%%%%%%%%%%%%%%%%%%%%%%%%%%%%%%
\begin{table*}[t!]
\centering
\caption{\label{tab:max_dress} Maximum dressing fractions $1-X_Z$ and $E_AT$ for nuclei from $^{56}_{26}$Fe to $^{231}_{91}$Pa propagating in a blackbody photon field of temperature $T$. $E_A$ is the energy of nucleus at which the maximum dressing occurs.}

\renewcommand{\arraystretch}{1.3}
\begin{tabular}{|c|c|c||c|c|c|}
\hline
Element & $E_AT~{\rm [eV^2]}$ & $\max(1 - X_{Z})$ & Element & $E_AT~{\rm [eV^2]}$ & $\max(1 - X_{Z})$ \\
\hline
$^{56}_{26}$Fe & $1.75 \times 10^{19}$ & 0.123 & $^{141}_{59}$Pr & $6.97 \times 10^{19}$ & 0.809 \\
\hline
$^{59}_{27}$Co & $2.06 \times 10^{19}$ & 0.140 & $^{144}_{60}$Nd & $7.56 \times 10^{19}$ & 0.821 \\
\hline
$^{59}_{28}$Ni & $2.23 \times 10^{19}$ & 0.157 & $^{145}_{61}$Pm & $7.56 \times 10^{19}$ & 0.831 \\
\hline
$^{64}_{29}$Cu & $2.42 \times 10^{19}$ & 0.179 & $^{150}_{62}$Sm & $8.21 \times 10^{19}$ & 0.841 \\
\hline
$^{65}_{30}$Zn & $2.63 \times 10^{19}$ & 0.198 & $^{152}_{63}$Eu & $8.90 \times 10^{19}$ & 0.850 \\
\hline
$^{70}_{31}$Ga & $2.85 \times 10^{19}$ & 0.223 & $^{157}_{64}$Gd & $9.66 \times 10^{19}$ & 0.858 \\
\hline
$^{73}_{32}$Ge & $3.09 \times 10^{19}$ & 0.247 & $^{159}_{65}$Tb & $1.14 \times 10^{20}$ & 0.870 \\
\hline
$^{75}_{33}$As & $3.35 \times 10^{19}$ & 0.269 & $^{164}_{66}$Dy & $1.23 \times 10^{20}$ & 0.881 \\
\hline
$^{79}_{34}$Se & $3.64 \times 10^{19}$ & 0.291 & $^{165}_{67}$Ho & $1.34 \times 10^{20}$ & 0.892 \\
\hline
$^{80}_{35}$Br & $3.64 \times 10^{19}$ & 0.319 & $^{167}_{68}$Er & $1.45 \times 10^{20}$ & 0.899 \\
\hline
$^{84}_{36}$Kr & $4.28 \times 10^{19}$ & 0.342 & $^{169}_{69}$Tm & $1.57 \times 10^{20}$ & 0.902 \\
\hline
$^{85}_{37}$Rb & $4.28 \times 10^{19}$ & 0.369 & $^{173}_{70}$Yb & $1.85 \times 10^{20}$ & 0.904 \\
\hline
$^{88}_{38}$Sr & $4.64 \times 10^{19}$ & 0.392 & $^{175}_{71}$Lu & $1.85 \times 10^{20}$ & 0.905 \\
\hline
$^{89}_{39}$Y & $4.64 \times 10^{19}$ & 0.416 & $^{178}_{72}$Hf & $2.01 \times 10^{20}$ & 0.907 \\
\hline
$^{91}_{40}$Zr & $4.64 \times 10^{19}$ & 0.441 & $^{181}_{73}$Ta & $2.36 \times 10^{20}$ & 0.908 \\
\hline
$^{93}_{41}$Nb & $5.46 \times 10^{19}$ & 0.463 & $^{184}_{74}$W & $2.36 \times 10^{20}$ & 0.909 \\
\hline
$^{98}_{42}$Mo & $5.93 \times 10^{19}$ & 0.485 & $^{187}_{75}$Re & $2.57 \times 10^{20}$ & 0.910 \\
\hline
$^{98}_{43}$Tc & $5.93 \times 10^{19}$ & 0.505 & $^{192}_{76}$Os & $2.78 \times 10^{20}$ & 0.912 \\
\hline
$^{101}_{44}$Ru & $5.93 \times 10^{19}$ & 0.531 & $^{193}_{77}$Ir & $3.02 \times 10^{20}$ & 0.913 \\
\hline
$^{103}_{45}$Rh & $5.93 \times 10^{19}$ & 0.551 & $^{195}_{78}$Pt & $3.27 \times 10^{20}$ & 0.914 \\
\hline
$^{106}_{46}$Pd & $5.93 \times 10^{19}$ & 0.570 & $^{197}_{79}$Au & $3.55 \times 10^{20}$ & 0.915 \\
\hline
$^{107}_{47}$Ag & $5.93 \times 10^{19}$ & 0.594 & $^{202}_{80}$Hg & $3.85 \times 10^{20}$ & 0.915 \\
\hline
$^{114}_{48}$Cd & $5.93 \times 10^{19}$ & 0.613 & $^{205}_{81}$Tl & $3.85 \times 10^{20}$ & 0.916 \\
\hline
$^{115}_{49}$In & $5.93 \times 10^{19}$ & 0.633 & $^{208}_{82}$Pb & $4.18 \times 10^{20}$ & 0.916 \\
\hline
$^{120}_{50}$Sn & $5.93 \times 10^{19}$ & 0.656 & $^{209}_{83}$Bi & $4.92 \times 10^{20}$ & 0.917 \\
\hline
$^{121}_{51}$Sb & $6.43 \times 10^{19}$ & 0.675 & $^{210}_{84}$Po & $4.92 \times 10^{20}$ & 0.917 \\
\hline
$^{126}_{52}$Te & $6.43 \times 10^{19}$ & 0.693 & $^{210}_{85}$At & $5.34 \times 10^{20}$ & 0.917 \\
\hline
$^{127}_{53}$I & $6.97 \times 10^{19}$ & 0.712 & $^{222}_{86}$Rn & $5.79 \times 10^{20}$ & 0.918 \\
\hline
$^{132}_{54}$Xe & $7.56 \times 10^{19}$ & 0.733 & $^{223}_{87}$Fr & $5.79 \times 10^{20}$ & 0.918 \\
\hline
$^{133}_{55}$Cs & $7.56 \times 10^{19}$ & 0.751 & $^{226}_{88}$Ra & $6.28 \times 10^{20}$ & 0.918 \\
\hline
$^{138}_{56}$Ba & $8.21 \times 10^{19}$ & 0.765 & $^{227}_{89}$Ac & $6.28 \times 10^{20}$ & 0.918 \\
\hline
$^{139}_{57}$La & $8.90 \times 10^{19}$ & 0.782 & $^{232}_{90}$Th & $6.28 \times 10^{20}$ & 0.919 \\
\hline
$^{140}_{58}$Ce & $9.66 \times 10^{19}$ & 0.797 & $^{231}_{91}$Pa & $6.28 \times 10^{20}$ & 0.919 \\
\hline
\end{tabular}
\end{table*}
%%%%%%%%%%%%%%%%%%%%%%%%%%%%%%%%%%
%%%%%%%%%%%%%%%%%%%%%%%%%%%%%%%%%%

\end{document}